# MULTIPLE-BUBBLE TESTING IN THE CRYPTOCURRENCY MARKET: A CASE STUDY OF BITCOIN

Sanaz Behzadi
Mahmonir Bayanati
Hamed Nozari[1]

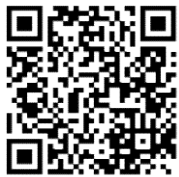
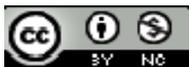

A B S T R A C T

*Economic periods and financial crises have highlighted the importance of evaluating financial markets to investors and researchers in recent decades. The asset price fluctuation is an innate quality of the market. However, fluctuations sometimes become abnormal and turn into unbridled spikes and sudden declines, causing irreversible damage to an economy and the trust of investors. In recent years, there have been severe price fluctuations in the cryptocurrency market. Some analysts have defined these fluctuations as price bubbles (i.e., economic bubbles). However, no specific definitions have yet been proposed for these price bubbles. The right-tail augmented Dickey–Fuller (RTADF) test was conducted on the time series data of 13 years (2008–2021) to analyze the presence of bubbles in the Bitcoin market. The RTADF results rejected the null hypothesis and confirmed the alternative hypothesis indicating the presence of bubbles in the Bitcoin market. In other words, there is some evidence of explosive behavior in a bubble. Moreover, the results of GSADF and SADF tests confirmed the presence of multiple bubbles in the Bitcoin market within 2008–2021.*

## 1. INTRODUCTION

Economic developments cause so many vicissitudes that economies sometimes peak and then plummet some other times (Auer, 2019). In this regard, the financial crises of recent decades have made asset markets become one of the hot topics discussed by economists and financial researchers (Bryan, 2020). The price fluctuations of an asset market can affect the other economic sectors (Chainalysis, 2021). There is extensive evidence of this phenomenon in various countries and markets throughout history (Biais et al., 2019). Generally, considering the time series of asset prices, we can infer that most of them have occasionally experienced severe fluctuations caused by national economic, social, and political events or global events such as international financial crises, oil shocks, sudden shifts in currency exchange policies, wars, and political instability. The outcomes of these events sometimes persist in an economy for a long time. As a result of such price fluctuations in a market, investors are encouraged to adjust their portfolios and rearrange their assets. Not only can this problem exacerbate confusion in the critical market, but it can also transfer fluctuations and bubbles to other markets (Kindleberger, 1991, Biais et al., 2018, Hatipoglu & Uyar 2020, Nneji, Brooks & Ward 2021).

In the history of economy, there have been some prominent cases of economic crises ensuing from bubbling increases in asset prices. Referred to as the Dutch Tulipmania, one of the early cases was the crisis caused by speculation in the Dutch tulip bulb market in the 17th century, when drastic spikes in the tulip price stimulated people to purchase tulip bulbs on credit in the hope of making a quick fortune. As a result, one tulip bulb was worth even more than many luxurious houses in Amsterdam. However, in 1637, the tulip bulb bubble burst, and many market activists went bankrupt (Bano et al., 2017). The Great Depression (1929–1939) in the US is another noteworthy exemplar of crises emerging from the growth and then the burst of bubbles. It has remained vivid in the memory of economists (Choi & Rocheteau, 2019). There are also some recent crises caused by bubbles in assets as follows: The Lost Decade in Japan in the late 1980s and the early 1990s; financial crises in several Latin American economies in the 1980s; financial crises in Southeastern Asian economies in the late 1990s; and many industrial economies in the recent international financial crisis (Nozari et al., 2023).

---

[1]Corresponding author: Hamed Nozari
Email: ham.nozari.eng@iauctb.ac.ir  83

In many cases, the asset price bubble phenomenon is accompanied by financial crises and will cause economies to incur huge costs. Therefore, it is essential to adopt appropriate methods for detecting the presence of price bubbles in financial assets to develop an initial warning system. The bubble detection problems must then be analyzed in detail (Auer, 2019).

The research rationale lies in the fact that it is complicated to correctly detect the price bubbles of assets due to uncertainty in the accurate identification of fundamental factors that are also stochastic. In particular, if there are innate bubbles that depend on foundations, the correct explanation of a bubble equation will be more difficult. Many of the existing methods for testing price bubbles have been criticized for various reasons. Therefore, this study uses novel methods for measuring price bubbles in the cryptocurrency market (Bitcoin) (Bonneau, Miller, Clark, Narayanan, Kroll & Felten 2015, Chiu & Koeppl 2017). Moreover, answering the question of whether the asset price increase is caused by fundamental economic factors or bubbles can help make different decisions and adopt diverse policies (Goren & Spiegelman, 2019). Furthermore, if the asset bubbles are left unchecked, they may lead to severe economic imbalances and decrease national revenues. Thus, policymakers must react to the formation of asset bubbles and be able to perceive the speculation transition pattern for implementing monetary policies to prevent the upcoming bubble overflows between markets and from one spot to another (Totonchi & Pourshalmani, 2021).

This study addresses the following questions:

Is there a significant relationship between the Bitcoin price bubble and its real value? Is there a significant relationship between the Bitcoin bubble profit and its real price? To answer these questions, a review of the research literature is first conducted in the area of financial bubbles. The research methodology is then introduced to test the collected data through the research model. The research results can be useful for investment portfolio optimization and monetary/financial policymaking.

## 2. PROBLEM STATEMENT AND RESEARCH BACKGROUND

Generally, there is no detailed consensus in the literature on the definition of a bubble and its causes (Radziwill, 2018). However, bubbles can be defined simply as steady massive increases in the price of an asset or a group of assets in a way that the initial price increase is caused by the expectations of upward trends in prices and the attraction of new buyers. This price increase is often accompanied by reverse expectations and steep declines in prices, a situation which bursts bubbles and then causes financial crises (Lamport et al., 2019).

Economists have tried to complete the definition of a bubble by considering the concept of fundamental factors, which are the economic factors that determine the prices of assets. Accordingly, a bubble is a part of a trend in an asset price that cannot be justified with fundamental factors (Goren & Spiegelman, 2019). Therefore, the prices of financial assets can be considered to consist of two components: a fundamental component and a bubble component. If the current price of an asset is high only because investors believe that it will increase in the foreseeable future without fundamental factors being able to explain such changes, there will then be price bubbles (Eyal et al., 2016).

Many of the studies testing the presence of price bubbles have been challenged for various reasons, e.g., low capability to detect bubbles, inability to determine when bubbles form and burst, inability to measure the magnitudes of bubbles, disregarding the possibility of negative bubbles, negligence of endogenous bubbles, and compatibility with only linear processes. In addition, the relationships of assets have been analyzed from different perspectives. However, previous studies have mainly analyzed the relationships of current prices in these markets or the contagion of their fluctuations spreading from one to another. If the presence of bubbles is considered possible, the price of an asset can be divided into a fundamental component and a bubble component. Hence, if the asset markets are related and if price bubbles are confirmed, are the relationships of these markets caused by the relationships of their fundamental factors? Do bubbles move from one market to another? Are the relationships of assets caused by both fundamental factors and bubbles? (Nozari et al., 2023).

Many studies have been conducted on price bubbles, and different tests have been proposed to discover bubbles. Shiller's variance bound test (Shiller 1981) and West's two-step test (1987) are two of the earliest methods for detecting stock price bubbles in the S&P 500 (the Standard and Poor's 500). The unit root test and the cointegration test are two of the most common methods for discovering asset price bubbles. They were proposed by Diba and Grossman (1984) and Campbell and Shiller (1988), respectively. These tests have empirically yielded contradictory results on the presence or absence of price bubbles. In the past two decades, these methods have been used extensively to discover asset price bubbles, especially those in the stock and housing markets. Criticizing the previous methods of bubble detection, Evans (1991) indicated that those tests were unable to detect an important class of bubbles.

According to Evans, the foregoing methods are unable to detect explosive bubbles when the candidate data face periodically collapsing bubbles. Hence, these tests fail to detect an important class of price bubble processes, i.e., busting and collapsing. In addition, these tests are true about linear processes; however, bubbles probably have nonlinear processes in the real world. Nevertheless, Evans failed to propose a model without such flaws. Charemza & Deadman (1995) criticized the cointegration test and the unit root test in order to analyze price bubbles. They concluded that those models were unable to search for and identify all bubbles. In other words, changes in the sample size, insufficiency of series data, and specific characteristics of series (e.g., severe fluctuations) may affect the capability of these tests to detect price bubbles. As a result of such criticisms, other methods have been proposed for bubble detection (Eyal et al., 2016).

Wu (1995) used Kalman filtering to estimate and test the currency exchange rate bubbles. Wu (1997) also employed Kalman filtering to analyze the stock market bubbles. Taylor and Peel (1998) conducted skewness and kurtosis tests to infer and prove the presence of price bubbles. According to their results, if the values of skewness and kurtosis indicate the non-normal distribution of a variable, the presence of a bubble is confirmed in the time series. Alexiou et al. (2019) conducted the flow test to discover and analyze rational bubbles in the housing market. Bierens (2005) used the Monte Carlo simulation to prove that the unit root test and the cointegration test would face explicit errors in nonlinear processes. Cunado et al. (2008) used the fractional integration test because they believed that the unit root test and the cointegration test were inefficient in rejecting the hypothesis indicating the presence of rational price bubbles. Chen et al. (2016) employed the conventional cointegration test and the threshold cointegration test to analyze the presence of rational price bubbles. Harman and Zuehlke 2004 analyzed price bubbles by using the hazard function (i.e., the duration dependence).

Philips et al. (2015) proposed the supremum augmented Dickey–Fuller (SADF) test based on the right-tail unit root tests to overcome the flaws of previous methods for detecting price bubbles, especially the Dicky–Fuller unit root test and the cointegration test. Their proposed method has the advantage of increasing the distinguishability of bubbles significantly in addition to testing the hypothesis of explosive behavior in prices. It also enables researchers to estimate the dates on which bubbles appear and vanish. Nevertheless, the SADF test faces a serious constraint, i.e., it is suitable for analyzing a single bubble period. However, multiple bubble periods may emerge in many time series. To overcome this constraint, Philips et al. (2015) generalized their methodology and proposed a generalized supremum augmented Dicky–Fuller (GSADF) test to identify multiple bubbles. Ever since, most of the studies have used this method to detect bubbles in financial asset markets.

The rational bubble models can be classified as two main categories: models for endogenous bubbles and models for exogenous bubbles. The endogenous models analyze bubbles regardless of changes in the fundamental values of assets. The variance limit test, unit root test, conventional cointegration test, and right-tail unit root test are the major exemplars of methods for detecting exogenous bubbles.

However, endogenous models consider the effects of changes in fundamental factors on the process of forming price bubbles. The endogenous bubble models were first introduced by Fruit and Obstfeld (1991), who proposed the foundation-dependent innate bubbles. These models consider the possibility of spreading shocks from fundamental factors to bubbles. Within the framework of innate rational bubbles, it is essential to determine how much of deviation in asset prices can be explained by a bubble component if the market foundations are exposed to continuous shocks. Kalman filtering can be considered the most important method of testing endogenous bubbles.

Zomorodian and Mahbubi (2022) analyzed the existence of long-term memories in four major cryptocurrencies (i.e., Bitcoin, Ethereum, Ripple, and Litecoin) from January 2016 to November 2019. They concluded that although the R/S outputs indicated the existence of long-term memories in all four cryptocurrencies, the model results and the ARFIMA outputs showed that Bitcoin and Ethereum had long-term memories. Therefore, previous prices can be taken into account to predict future prices, a finding which rejects the hypothesis of efficient markets and confirms the existence of speculative motives for these cryptocurrencies. In a study entitled Identifying Multiple Bubbles in the Tehran Stock Exchange through Right-Tail Markov–Switching Unit Root Test, Hosseinzadeh (2021) concluded that the price/earnings ratio (P/E) had depicted explosive behavior in several intervals since 2001. In other words, P/E experienced bubble conditions. According to the results of analyzing smoothed probabilities for bubble intervals and other variables (e.g., the total nominal index, the total real index, P/E, and the ratio of the market value to the GDP), the most substantial bubble in the history of the Tehran Stock Exchange appeared from March 2020 to June 2020.

Abolhasani and Samadi (2020) used the time series analysis to study the price determinants of Bitcoin and Ethereum within 2014–2020 and those of Ethereum within 2016–2020 through OLS and VECM. According to their results, the global price of gold was the most effective factor determining the prices of Bitcoin and Ethereum. At the same time, the dollar and euro exchange rates and the global price of gold had negative relationships with the values of cryptocurrencies in the short run.

In a study entitled Bubble Testing in the Bitcoin Market, Hoang and Morken (2018) concluded that six substantial bubbles were observed within 2011–2018 and spanned 24 to 123 days. According to their statistical evidence, there are no bubbles in the Bitcoin market now. They also realized that those bubbles might not have any information on logical expectations but might have some information on illogical emotions. This finding is consistent with the theory proposed by Google Trends, RSI, and the "bubble stages" model.

Philips et al. (2015) proposed a method for dating bubbles and determining their types in terms of being single or multiple by employing RADF, SADF, and GSADF techniques. In fact, previous methods of bubble detection (e.g., skewness, kurtosis, and infinite regress) were only able to detect bubbles but were unable to date bubbles. However, the novel method proposed by Philips et al. solved this problem. In addition to dating bubbles, this method determines the types of bubbles in terms of being singular or multiple.

This study tests multiple bubbles in two steps. Firstly, a novel method is proposed to test the multiple-bubble hypothesis in the Bitcoin market and the presence of numerous bubbles within one interval in each of the designated cryptocurrency markets of Iran (Rasekhi, Shahrazi & Alami 2020).. Secondly, the fundamental components and bubble components of these markets are distinguished. The multiple-bubble hypothesis in an economy and the overflow of bubbles between cryptocurrency markets and the Bitcoin market are then analyzed. Hence, this study aims to analyze multiple-bubble testing in the cryptocurrency markets of Bitcoin.

## 3. METHODOLOGY

In this applied retrospective study, the previous financial data of cryptocurrency markets are used for hypothesis testing. The research results can directly be employed by users to make decisions. This research is also an analytical study in which the hypotheses are tested by using the resultant data and introducing some hypotheses. Moreover, the panel data regression model is utilized to discover the correlation between two variables. The research information included the data of financial reports on cryptocurrency markets extracted from relevant information banks. The data were then processed to estimate the research models and provide a basis for hypothesis testing. The necessary raw data of markets were collected from https://tsetmc.com/ via Rahavard Novin for hypothesis testing. The data were then compared to remove various mismatches. The results were transferred to Excel worksheets, and the final analysis was conducted in EViews 10.

Research Hypotheses:

First Hypothesis: There is a significant relationship between the Bitcoin price bubble and its real value.

Second Hypothesis: There is a significant relationship between the Bitcoin bubble profit and its real price.

## 4. RESEARCH FINDINGS

### 4.1 Bubble Detection

In Iranian studies, various tests, e.g., infinite regress, skewness, kurtosis, cointegration, fractional integration, and unit root test, were used for bubble detection. However, the foregoing tests are unable to determine when bubbles occur; they are only able to determine the presence or absence of bubbles. The RTADF-based tests must be conducted to determine the dates at which bubbles occur. In this study, four Dickey–Fuller tests are first employed for bubble detection: the right-tail augmented Dickey–Fuller test, the rolling window ADF, the supremum ADF (SADF), and the generalized SADF (GSADF). In each test, the rejection of the null hypothesis indicates the presence of an asset price bubble. In the second stage, RADF and GSADF tests are conducted to determine the dates at which bubbles occur. The resultant values were acquired after 10,000 iterations.

Table (1) lists the bubble detection tests with respect to the right-tail standard test skewness, in which the null hypothesis is based on the unit root, whereas the opposite hypothesis is based on the presence of bubbles. Generally, the results of four tests reject the hypothesis stating the presence of a unit root. In other words, the results did not reject the presence of bubbles in the Bitcoin price within the research period.

**Table 1**. Bubble detection tests (Reference: Research Calculations)

| Statistic Index | ADF | RADF | SADF | GSADF |
|---|---|---|---|---|
| Bitcoin | 4.06 (0.000) | 4.69 (0.000) | 4.06 (0.000) | 5.78 (0.000) |

*The numbers between parentheses indicate the P-values.

### 4.2 Dates of Bubbles

Based on the proposed methodology, this subsection determines the times at which bubbles emerge, explode, and burst completely. In a period that includes multiple bubbles, the explosion time corresponds to the largest bubble of that period. In all of the following charts, the upper curve (green) indicates the index of interest, whereas the middle curve (red) represents critical values at 95%. Finally, the lower curve (blue) represents the statistics of ADF, rolling window ADF, and GSADF tests. In such tests, the test statistic is determined for dating based on the real price chart. The curve of critical values is then drawn for decision-making. Now if the test statistic exceeds the predetermined critical value, there will be a bubble. In this case, the first time when the blue curve exceeds critical values marks the beginning of a bubble area, and the time when it reaches below critical values again marks the date at which the bubble starts to disappear completely. Furthermore, the time when the blue curve reaches its climax is considered the bubble explosion time, whereas the complete disappearance refers to the situation in which the blue curve crosses the critical value (red line) and stands below it. The bubble area includes the point at which it appears until the point at which it disappears completely.

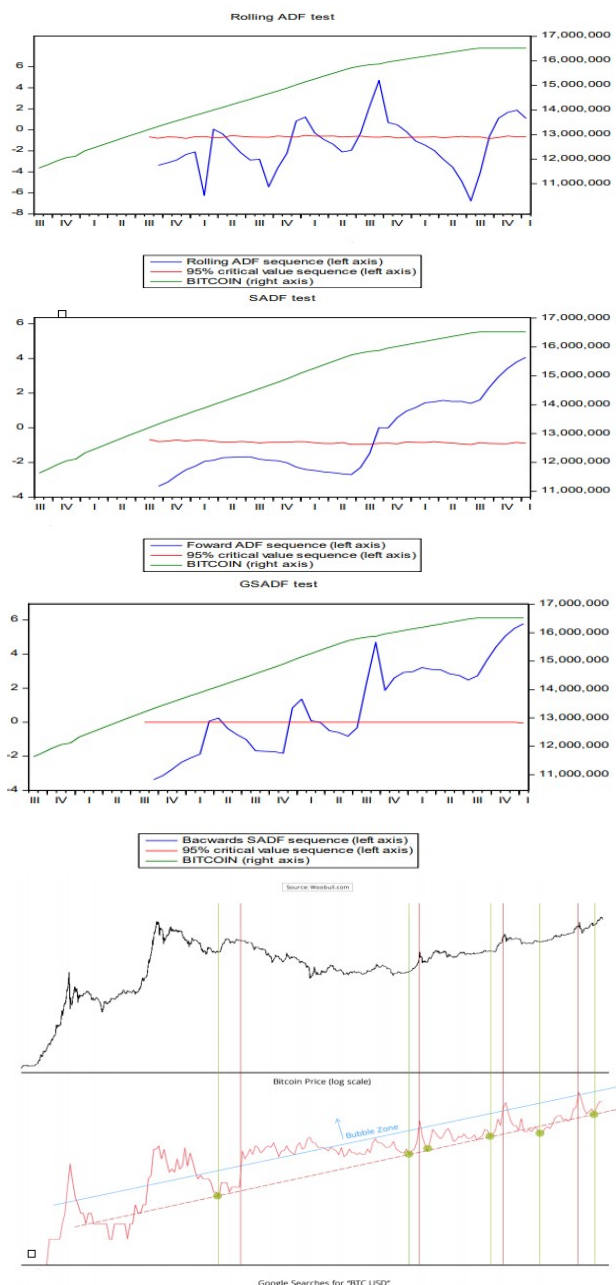

**Figure 1**. Estimation Charts

According to the test results, the following conclusions are drawn on the Bitcoin bubble:

- According to the RADF test results, there were four bubble periods with a single structure. There were bubbles in 46% of the period, whereas there were no bubbles in 54% of the period. The longest bubble period concerned the third bubble lasting for 7 months (September 23, 2014 to March 20, 2015), whereas the shortest non-bubble period lasted for two months (between 2012 and 2013).
- According to the SADF test results, there was a continuous long-term bubble period with a single structure. There were bubbles in 41% of the period, whereas there were no bubbles in 59% of the period. The long-term bubble period lasted for nearly 17 months (March 21, 2019 to September 22, 2019) with an upward trend.
- According to the GSADF test results, there were three bubble periods, two of which had single structures, whereas the third one had multiple structures. There were bubbles in 61% of the period, whereas there were no bubbles in 39% of the period. The longest bubble period concerned the third bubble that lasted for 18 months (2019–2021).

### 4.3 Research Test Results

Table (2) reports the results of the RTADF tests:

**Table 2**. The results of RTADF tests

| Test | P-Value | Statistic | Critical Value of 99% Reliance | Critical Value of 95% Reliance | Critical Value of 90% Reliance |
|---|---|---|---|---|---|
| ADF | 0.007 | 1.021 | 0.9550 | 0.010 | -0.389 |
| SADF | 0.000 | 2.710 | 1.817 | 1.433 | 1.141 |
| GSADF | 0.000 | 3.453 | 2.566 | 2.076 | 1.772 |

The statistics of RTADF are calculated concerning the backward multiple regressions in which there are different numbers of observations and initial observation for each regression. Within this framework, the SADF test statistic is calculated for each regression based on the number of observations to determine the dates at which each separate bubble appears and bursts. Moreover, the GSADF test statistic is employed to determine the occurrence of at least one bubble in the entire sample. The resultant values of each statistic are then compared with appropriate time series of critical values.

Table (2) reports the RTADF test results, rejecting the null hypothesis and confirming the alternative hypothesis that indicates the presence of bubbles in the Bitcoin market. In other words, there is some evidence of explosive behavior in a bubble. Moreover, GSADF and SADF tests confirmed the presence of multiple bubbles in Iran's Bitcoin market within 2009–2021. Given the relationship between financial bubbles and the occurrence of crises, the presence of bubbles within the foregoing period can be one of the factors that intensified instability in Iran's economy in recent years. These tests helped determine the dates at which each bubble appeared and burst.

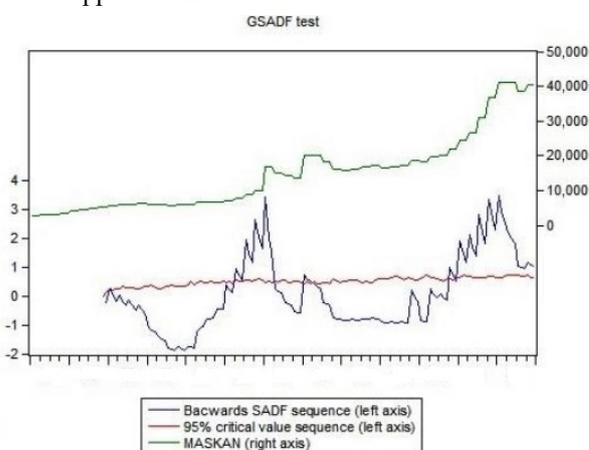

**Figure 2**. The GSADF test for detecting multiple bubbles

Figure (2) reports the GSADF test results. Accordingly, the Bitcoin price index experienced a bubble from September 23, 2014 to March 20, 2015; however, this bubble burst after a while in April 2015. Within 2011–2021, a price bubble was observed from March 20, 2020 to September 21, 2020. This bubble approached the brink of elimination from March 21, 2021 to September 22, 2021 when the presidential election was held in Iran. In conclusion, a considerable part of the dramatic spike in the Bitcoin price was caused by the presence of a price bubble.

According to the results, SADF and GSADF tests are clearly characterized by the fact that they can detect multiple bubbles and determine the dates at which bubbles appear and burst when we face the periodic collapsing of bubbles. However, the conventional left-tail unit root tests lack these capabilities. The null hypothesis of the left-tail augmented Dickey–Fuller test and the Phillips–Perron (PP) test indicated the non-stationarity and presence of price bubbles in the Bitcoin market, whereas the alternative hypothesis indicated the stationarity and absence of price bubbles in the Bitcoin market. However, the alternative hypothesis of the RTADF unit root test indicated explosive behavior in the Bitcoin price index bubble. Table (3) reports the results of the conventional left-tail augmented Dickey–Fuller unit root tests and the Philips–Perron (PP) test for the Bitcoin price index within 2009–2021.

**Table 3**. The results of LTADF tests

| Test | P-Value | Statistic | Critical Value of 99% Reliance | Critical Value of 95% Reliance | Critical Value of 90% Reliance |
|---|---|---|---|---|---|
| ADF | 0.9967 | 1.02156 | -3.472 | -2.880 | -2.576 |
| SADF | 0.9965 | 1.00445 | -3.472 | -2.880 | -2.576 |

According to Table 3, the statistics and P-values of the conventional ADF unit root tests and the Philips–Perron (PP) test did not reject the null hypothesis indicating the stationarity and presence of unit root tests and bubble prices in the Bitcoin market within the research period. However, the left-tail unit root tests determine only the presence of absence of bubbles within an entire period and fail to detect multiple bubbles or the periodical collapsing of bubbles. Nevertheless, the right-tail GSADF unit root test evaluates the explosive behavior of a variable to properly detect the presence of multiple bubbles within a period and determine the dates at which each bubble emerges or bursts.

## 5. CONCLUSION

In recent years, there have been relatively substantial fluctuations in the prices of cryptocurrencies, especially that of Bitcoin. These fluctuations are considered price bubbles by some researchers; however, they have not provided any specific detailed definitions for these price bubbles. In this study, an econometric RTADF-based method was employed to analyze the presence of bubbles in the Bitcoin market. For this purpose, the time series data of 13 years (2008–2021) were used.

The RTADF test results rejected the null hypothesis but confirmed the alternative hypothesis indicating the presence of bubbles in the Bitcoin market. In other words, there was some evidence of explosive behavior in a bubble. Moreover, the results of GSADF and SADF tests confirmed the presence of multiple bubbles in the Bitcoin market within 2008–2021. Given the relationship between financial bubbles and the occurrence of crises, the presence of bubbles within the foregoing period can be one of the factors that intensified instability in Iran's economy in recent years. These tests helped determine the dates at which each bubble appeared and burst. According to the GSADF test results, the Bitcoin price index experienced a bubble from September 23, 2014 to March 20, 2015; however, this bubble burst after a while in April 2015. Within 2011–2021, a price bubble was observed from March 20, 2020 to September 21, 2020. This bubble approached the brink of elimination from March 21, 2021 to September 22, 2021 when the presidential election was held in Iran. In conclusion, a considerable part of the dramatic spike in the Bitcoin price was caused by the presence of a price bubble.

According to the test results, the following conclusions are drawn on the Bitcoin bubble:

- According to the RADF test results, there were four bubble periods with a single structure. There were bubbles in 46% of the period, whereas there were no bubbles in 54% of the period. The longest bubble period concerned the third bubble lasting for 7 months (September 23, 2014 to March 20, 2015), whereas the shortest non-bubble period lasted for two months (between 2012 and 2013).
- According to the SADF test results, there was a continuous long-term bubble period with a single structure. There were bubbles in 41% of the period, whereas there were no bubbles in 59% of the period. The long-term bubble period lasted for nearly 17 months (March 21, 2019 to September 22, 2019) with an upward trend.
- According to the GSADF test results, there were three bubble periods, two of which had single structures, whereas the third one had multiple structures. There were bubbles in 61% of the period, whereas there were no bubbles in 39% of the period. The longest bubble period concerned the third bubble that lasted for 18 months (2019–2021).

Bitcoin is the first ever cryptocurrency created worldwide. It was first introduced globally in an article published by a programmer named Satoshi Nakamoto. Bitcoin is a digital cryptocurrency that has attracted many investors in recent years. According to the results of analyzing the change trend in the price of this cryptocurrency, there have been substantial fluctuations in recent years. This finding strengthens the hypothesis that states there are bubbles in this market. Therefore, since no Iranian studies have analyzed the Bitcoin market, this study aimed to use novel SADF, GSADF, and RADF tests proposed by Philips et al. (2015) to analyze the bubbles emerging in this market.

According to the RADF test results, there were four bubble periods with a single structure. There were bubbles in 46% of the period, whereas there were no bubbles in 54% of the period. According to the SADF test results, there was a continuous long-term bubble period with a single structure. In this test, there were bubbles in 41% of the period, whereas there were no bubbles in 56% of the period. Interestingly, this bubble has not yet exploded and has experienced an upward trend so far. The GSADF test results indicated that there were three bubble periods, two of which had single structures, whereas the third period had multiple structures. Furthermore, there were bubbles in 61% of the period, whereas there were no bubbles in 39% of the period. According to the results, the bubble periods determined by RADF and GSADF tests were nearly the same. Therefore, the Bitcoin market had bubbles within 2019–2021, and the final bubble has persisted and has not yet exploded.

Different investigations indicate that Bitcoin has followed the characteristics of an asset rather than those of a currency. Therefore, it is the main factor that causes investors to show speculative behaviors in asset markets. Moreover, the results of analyzing the bubbles of Bitcoin indicated that they were irrational, something which is somehow caused by following existing excitements in different asset markets such as the Bitcoin market. The presence of bubbles in financial markets can destabilize the economy of a country. According to the results of this study, the outcomes of this crypto currency have penetrated Iran's financial market and may lead to substantial fluctuations in financial markets. Therefore, researchers are recommended to conduct a separate study to analyze the possibility of bubble transition between the Bitcoin market and other financial markets such as the stock market and the gold market to determine means of communication between financial markets. Moreover, investors are recommended to make decisions on the arrival of Bitcoin in the stock market, concerning its arrival and that of other crypto currencies in financial markets and the transition of bubbles from the crypto currency market to other financial markets, e.g., the stock market.

**Sanaz Behzadi**
Faculty of Technology and Industrial Management, Department of Management, West Tehran Branch, Islamic Azad University, Tehran, Iran
Sanaz.behzadi72@gmail.com
**ORCID:** 0009-0002-0513-0894

**Mahmonir Bayanati**
Faculty of Technology and Industrial Management, Department of Management, West Tehran Branch, Islamic Azad University, Tehran, Iran
bayanati.mahmonir@wtiau.ac.ir
**ORCID:** 0000-0002-4236-6110

**Hamed Nozari**
Department of Management, Azad University of the Emirates, Dubai, UAE
Ham.nozari.eng@iauctb.ac.ir
ORCID0000-0002-6500-6708